\documentclass{aastex631}

\usepackage[bottom]{footmisc}
\usepackage[normalem]{ulem}

\usepackage{lineno}
\linenumbers

\begin{document}
\nolinenumbers
\title{Distinguishing between flaring and non-flaring solar active regions with height-dependent polarity inversion line lengths \footnote{resubmitted on 19/10/2026}}
\author[0000-0002-1367-8326]{Shreeyesh Biswal}
\affiliation{Solar Physics \& Space Plasma Research Center (SP2RC), School of Mathematical and Physical Sciences, University of Sheffield, Hounsfield Road, Sheffield, S3 7RH, UK}
\affiliation{Department of Physics - Section Astrogeophysics, University of Ioannina, Ioannina - 45110, Greece}

\author[0000-0002-0049-4798]{Marianna B. Kors\'os}
\affil{University of Sheffield, School of Electrical and Electronic Engineering, Amy Johnson Building, Portabello Street, Sheffield, S1 3JD, UK}
\affiliation{Department of Astronomy, E\"otv\"os Lor\'and University, P\'azm\'any P\'eter s\'et\'any 1/A, Budapest, H-1117, Hungary}
\affiliation{Gyula Bay Zoltan Solar Observatory (GSO), Hungarian Solar Physics Foundation (HSPF), Pet\H{o}fi t\'er 3., Gyula, H-5700, Hungary}

\author[0000-0001-6913-1330]
{Manolis K. Georgoulis}
\affiliation{Johns Hopkins University Applied Physics Laboratory Laurel, MD 20375, USA}

\author[0000-0003-0475-2886]{Alexander Nindos}
\affiliation{Department of Physics - Section Astrogeophysics, University of Ioannina, Ioannina - 45110, Greece}

\author[0000-0003-3345-9697]
{Spiros Patsourakos}
\affiliation{Department of Physics - Section Astrogeophysics, University of Ioannina, Ioannina - 45110, Greece}

\author[0000-0003-3439-4127]{Robertus Erd\'elyi}
\affiliation{Solar Physics \& Space Plasma Research Center (SP2RC), School of Mathematical and Physical Sciences, University of Sheffield, Hounsfield Road, Sheffield, S3 7RH, UK}
\affiliation{Department of Astronomy, E\"otv\"os Lor\'and University, P\'azm\'any P\'eter s\'et\'any 1/A, Budapest H-1117, Hungary}
\affiliation{Gyula Bay Zoltan Solar Observatory (GSO), Hungarian Solar Physics Foundation (HSPF), Pet\H{o}fi t\'er 3., Gyula H-5700, Hungary}


\phantom{...}

\begin{abstract}
\phantom{...}

Polarity Inversion Lines (PILs) are regions on the Sun’s surface across which the vertical magnetic field changes polarity. This study examines whether the total length of PILs within a height range of 0-3.24 Mm above the photosphere can distinguish between strongly flaring active regions, with at least one flare of GOES (Geostationary Operational Environmental Satellite) class M1 or higher, and non-flaring active regions, where the strongest flare remains weaker than GOES class C1. Using magnetic field extrapolations on a dataset of 40 active regions, the study compares the 48-hr evolution of the total PIL length prior to the first major flare in strongly flaring regions with identical time intervals in non-flaring regions. It is seen that the 48-hr maximum of the total PIL length differs clearly between the two groups. This distinction becomes stronger at heights about 1-2 Mm above the photosphere. It is further seen that at about 1.8 Mm, the maximum value of the total PIL length, for around (i) 78 percent of weakly flaring regions (where the strongest flare is a C-class flare) and (ii) 96 percent of non-flaring active regions, is less than the minimum value of the total PIL length of their strongly flaring counterparts. These results suggest that strongly flaring active regions may be distinguishable from non-flaring or weakly flaring ones using thresholds on PIL lengths in the lower solar atmosphere. Further, this study highlights the importance of studying chromospheric magnetic fields in flare precursor investigations. 

\end{abstract}
\vspace{-1.0cm}

\section{Introduction}
\label{sec:intro}

\phantom{...}

Solar energetic events, such as flares \citep{Kopp1976} and coronal mass ejections (CMEs) (\citealp{Low1994}; \citealp{Dere1999}) are explosive phenomena driven by magnetic instabilities in the Sun's atmosphere. Since these phenomena release vast amounts of energy and energetic particles into the inner heliosphere, they significantly impact electronic, telecommunication and satellite-based infrastructure on Earth and geospace, often commonly referred to as the technosphere. Hence, forecasting these energetic phenomena is of key importance to the space weather research community in order to secure our critical infrastructure in the technosphere. \\

Forecasting space weather often involves studying the morphology of solar magnetic fields in the only place they can be measured routinely currently, namely, the solar photosphere. Such observations are available from ground- and space-based magnetographs. This is achieved by analysing and quantifying key parameters of solar magnetograms, that are known as proxies or predictors, characteristic to regions where the magnetic field is complex and disturbed significantly from its minimum-energy state. One such feature is the polarity inversion (i.e., neutral) lines (PILs) in the photosphere and lower solar atmosphere (LSA), which demarcate magnetic flux patches of opposite polarities. The strongest PILs appear in magnetic flux accumulations known as active regions (ARs). While commonly called 'neutral lines' to indicate magnetic neutrality in a form of a zero vertical or radial field, strong (i.e., flux-massive) PILs are also characterised by areas of high-shear (\citealp{Zirin1993}; see examples in \citealp{Yang2004}; \citealp{Deng2006}) in their (strongly non-zero) horizontal field. PILs are also closely linked with solar eruptive phenomena (see \citealp{vanDriel-Gesztelyi2015} and references therein). Numerical models for CME initiation also confirm that ARs with strong PILs are invariably associated with high velocity shear flow (\citealp{Amari2003, Roussev2004}). These plasma flows possibly occur as a response to the Lorentz force acting along PILs in the solar atmosphere \citep{Manchester2008}. Further, recent studies have shown that shearing arising out of collision between emerging non-conjugated bipoles in the lower solar atmosphere allows for the formation of compact PILs, leading to shearing and flux cancellation, and consequently solar flares and CMEs \citep{Chintzoglou_2019}. It has been shown that both the vertical component \citep{Sudol2005} and the horizontal component (\citealp{Wang1992, Wang_2019, Wang2006}) of the photopsheric magnetic field undergo rapid irreversible changes at the time of flare occurrence. Using MHD simulations, \cite{Bian2023} verified that especially close to the PILs, the horizontal magnetic field increases significantly over a scale of several minutes after flare occurrence. \\

Having outlined the key characteristics of PILs, it is clear that detection algorithms must account for both the vertical and horizontal magnetic field variations across PILs, in order to properly identify those associated with strong shear (arising from enhanced horizontal fields) and significant vertical flux gradients. In this connection, there exist a number of predictors, either linked to PILs or proxies of PILs themselves, in the literature that have been studied in the past. For example, from a study of PILs, \citet{Falconer2003} found the strong shear length $L_{SS}$ (computed from vector magnetograms) and the strong gradient length $L_{SG}$ (a proxy for $L_{SS}$ computed from LOS magnetograms) to be good indicators for non-potentiality, and thus suitable candidates for the prediction of CMEs. In this context, non-potentiality refers to the presence of free energy in the ARs arising due to stressing, shearing or twisting of field lines. Similarly, \citet{Schrijver2007} introduced a parameter called the $R$-value, that is an indicator of the flux around PILs that have a strong horizontal gradient in the vertical magnetic flux. \citet{Mason2010} introduced a parameter strongly correlated to $L_{SG}$ called $GWILL$ that is computed as a line integral of the gradient in the vertical magnetic field along the primary/longest PIL.  Based on the study of photospheric flow fields, \citet{Park2018} showed that the line integral of the shear flow speed (along PIL segments associated with (i) a strong horizontal magnetic field and (ii) a strong gradient of the vertical magnetic field) is a good precursor for strong flares. \\

Traditionally, flare precursors and predictors have been studied on the photosphere. However, recently it has been shown that studying flare precursors in a certain height range within the LSA, called the 'optimal height range' (OHR), could help in the increase of the lead time before strong flares, especially GOES X-class. For example, \citet{Korsos2020} found that it is possible to improve the lead time by 2-8 hr if the $WG_M$ morphological parameter \citep{Korsos2015} is studied between 1-1.8 Mm above the photosphere. From investigations of the emergence component of the helicity flux within 0.36-1.5 Mm above the photosphere, \citet{Korsos2022} demonstrated that the lead time may be improved by 2-8 hr. Similarly, \citet{Biswal2024} showed that a jump in $R$-value could be observed between 48-68 hr before the first X-class flare for emerging ARs within a height range of 0.36-3.24 Mm above the photosphere. Motivated by such results, our main objective here is to investigate whether the PIL length in similar height ranges (above the photosphere), obtained by the extrapolation of the photospheric magnetic field, could help distinguish between flaring and non-flaring ARs. This action could potentially give valuable insights on how magnetic field configuration differs between flaring/eruptive and non-flaring, or weakly flaring, ARs. \\

Since longer PILs usually indicate more complex magnetic field structures \citep{Mason2010}, we choose to study the time series data for PILs in an AR using two parameters: (i) $L_{tot}$ - denoting the total length of PILs observed at any specific point in time, within a given time series and (ii) $L_{tot}^*$ - denoting the maximum value of $L_{tot}$ within the time series. While $L_{tot}$ is useful to trace the overall evolution pattern of PIL length, $L_{tot}^*$ gives an estimate of the PIL length in the AR's most complex configuration. In this context, details of the $L_{tot}$ time series dataset for different AR categories are discussed in Section \ref{sec:ardataset}. This is followed by (i) a brief discussion of the extrapolation methodology and (ii) the algorithm used to compute $L_{tot}$ in the extrapolated data in Section \ref{sec:methods}. A few examples of PIL detection and $L_{tot}$ evolution patterns at different heights are presented in Section \ref{ssec:vis}. Further results on how an OHR (to separate flaring ARs from non-flaring ARs) may be determined using $L_{tot}^*$ are discussed in Section \ref{ssec:attenuation}. Sections \ref{ssec:samp} and \ref{ssec:pop} then identify the most optimal height within the OHR (determined in Section \ref{ssec:attenuation}) using descriptive and inferential statistical approaches, respectively. The overall implication of the study on flare predictability and forecasting, along with the concluding remarks are presented in Section \ref{sec:conc}. \\

\section{ACTIVE REGION DATASET} 
\label{sec:ardataset}

\phantom{...}

Our dataset see Appendix \ref{app:dataset}) consists of 40 ARs belonging to solar cycle 24, out of which 24 ARs produced flare(s) with strength $\ge$ C1.0. These 24 flaring ARs are further classified as (i) X-class flaring ARs or (ii) M-class flaring ARs or (iii) CFARs (C-class flaring active regions), depending on the GOES flare class of the strongest flare produced by the AR (see Table \ref{table:dataset1} in Appendix \ref{app:dataset}). For each of the three flaring groups, we study $L_{tot}$ for a 48 hr interval immediately before the first flare (belonging to the same class as the strongest flare) within a height range of 0-3.24 Mm (see details in Section \ref{sec:methods}). For example, AR 12673 is considered as an X-class flaring AR as the strongest flare was an X9.3 flare (2017/09/06 11:53 UTC). However, because the first X-class flare for AR 12673 was an X2.2 flare (2017/09/06 08:57 UTC), we study $L_{tot}$ in the 48 hr interval immediately preceding the (earlier) X2.2 flare and also take a note of the $L_{tot}^*$ value in that interval. Since it is known that ARs that have flared previously are more likely to flare again (\citealp{Zirin1991}; \citealp{Park2020}), focusing on the pre-flare evolution prior to the first flare ensures fairness in the analysis. For simplicity (see Section \ref{ssec:attenuation}), we consider the first two flaring groups as  strongly flaring active regions (denoted as SFARs). \\

Our dataset consists of 16 non-flaring active regions (denoted as NFARs henceforth), i.e. ARs not producing any flare with strength $\ge$ C1.0. Unlike flaring active regions (X/M/C-class), NFARs do not have a reference time point at which an analysis may be made to compare them with flaring ARs. To explore all relevant possibilities within the transit time of NFARs through the visible solar disk, we sampled the available data into as many distinct, non-overlapping  intervals as possible, following which we have 45 independent and distinct 48-hr time intervals (see Table \ref{table:dataset2}  in Appendix \ref{app:dataset}). \\

Since CFARs are often considered weakly flaring ARs in the literature (see for e.g. \citealp{Wang2022}; \citealp{Liu2023}), we also study them like NFARs by sampling the data independently, ignoring the time of occurrence of the first C-class flare. By doing so we follow the behaviour of $L_{tot}^*$ in CFARs beyond the occurrence of the first C-class flare to explore the possibility of a difference in the $L_{tot}^*$ evolution between CFARs in general, and the pre-flaring PIL lengths in SFARs. Hence we also sampled the available data for 8 CFARs into 23 independent and distinct 48-hr time intervals, irrespective of the time of occurrence of the first C-class flare (see Table \ref{table:dataset2} in Appendix \ref{app:dataset}). For example, if we take the case of AR 11730, the first C-class flare, i.e. the C5.9 flare,  occurred on 2013/04/29 (20:22 UTC). The 48-hr interval immediately before this flare is studied within the CFAR data (see Table \ref{table:dataset1} in Appendix \ref{app:dataset}). However, when we independently sample the available  magnetogram data for this AR, we obtain 3 distinct and sequential 48-hr time series (see Table \ref{table:dataset2} in Appendix \ref{app:dataset}) between Apr 26 - May 2, 2013 (00:00 UTC). \\

\section{ANALYSIS METHOD}
\label{sec:methods}

\phantom{...}

\subsection{Extrapolation of input data}
\label{ssec:ext}

\phantom{...}

Since our objective is to find the magnetic field configuration above the photosphere in ARs and subsequently compute $L_{tot}$ from the reconstructed field, our first step in this regard is to perform a Potential-Field (PF) extrapolation, based on the Fast Fourier Transform approach (\citealp{alissandrakis81}; \citealp{Gary1989}) on each AR magnetogram. The input vector magnetogram data is the radial component of the magnetic field ($B_r$), taken from the Helioseismic and Magnetic Imager (HMI), a magnetograph onboard the Solar Dynamics Observatory (SDO) mission (\citealp{Pesnell2012}; \citealp{Scherrer2012}). The data product we use is either the Spaceweather HMI Active Region Patch (SHARP) data or a crop-out of it, in case the original data contains more than one NOAA-numbered AR (\citealp{Bobra2014}). The PF extrapolation generates a 3D vector field solution from the $B_r$ of the AR, as observed in the photosphere. In particular, this analysis is restricted to the LSA and uses 10 2D $B_z$ maps produced by PF extrapolation which are equally spaced at steps of 0.36 Mm ($\sim$0.5\arcsec) up to 3.24 Mm, similar to the approach of \cite{Biswal2024}. The PF extrapolation is performed at intervals of 1 hr for the 48 hr observation time before the flare, generating a time series of magnetic field maps at each height. During the observation time, all ARs in our dataset were located between 60$^\circ$ EW of the central meridian to avoid severe projection effects \citep{Bobra2014}. This ensures that $L_{tot}$ calculations are based on reliable measures of the vertical magnetic field. While there are more elaborate extrapolation methods such as Non-Linear Force Free Field (NLFFF) method \citep{Wiegelmann2012}, we chose to use the PF method because it is sufficient enough for a quick and computationally inexpensive calculation of the 3D magnetic field, at the same time preserving the main characteristics of the configuration (\citealp{Korsos2024}).\\ 

\subsection{Computation and analysis of PIL parameters}
\label{ssec:comp}

\phantom{...}

To identify PIL(s), one way is to apply PIL masks of suitable field thresholds and appropriate convolution based image smoothing filters in order to highlight magnetic flux patches of opposite polarities and the PIL(s) separating them (\citealp{Schrijver2007}; \citealp{Wang_2019}; \citealp{Sun_2021}; similar approach in \citealp{Ji_2023}). \cite{Mason2010} followed a different approach where the input magnetogram is first smoothed and subsequently the vector potential field is computed for the identification of zero Gauss contours (\citealp{alissandrakis81}, \citealp{Falconer2002}). When these contours are identified, they must be above a certain horizontal field gradient threshold, so that PILs are considered only from strong shearing regions. The procedure is carried out at different smoothing thresholds and the outputs in each case are superposed to find overlapping regions, or the PILs as we understand them (refer to Figure 2 in \citealp{Mason2010}). Another interesting approach for detecting PILs in an AR involves the use of a computer vision based image segmentation algorithm \citep{Sadykov_2017}. Given the wide variety of algorithms and techniques employed to detect PILs in the literature (\citealp{Falconer2002}; \citealp{Falconer2003}; \citealp{Schrijver2007}; \citealp{Mason2010}; \citealp{Sadykov_2017}; \citealp{Sun_2021}; \citealp{Wang_2019}, \citealp{Ji_2023}), we selected an approach that incorporates physics-based control parameters computed from both the vertical and horizontal components of the magnetic field. Hence our method of detecting PILs is adapted from the FLARECAST project\footnote{{\url{https://dev.flarecast.eu/stash/projects/FE}}} \citep{Georgoulis2021}, which is largely modelled along the detection methodology provided by \citet{Mason2010}. In this approach, the control parameters are (i) $B_h$, i.e. the horizontal magnetic field, and (ii) $B_{zg}$, i.e. the spatially averaged vertical magnetic field computed within a 10 $\times$ 10 pixel box (3.5 $\times$ 3.5 Mm). The latter is evaluated separately for the positive and negative masks defined as $B_z^+ = \max(B_z,0)$ and $B_z^- = \min(B_z,0)$. Two thresholds are applied to $B_h$ - (i) $B_{h1} > 120$ G for an initial detection and (ii) $B_{h2} > 100$ G for a finer detection. Likewise, $B_{zg}$ uses two thresholds - for $B_z^+$, (i) $B_{zg1} > 20$ G and (ii) $B_{zg2} > 8$ G; for $B_z^-$, (i) $B_{zg1} < -20$ G and (ii) $B_{zg2} < -8$ G, corresponding to the coarser and finer detections, respectively. The superposition of coarser and finer detections provides a list of all PIL fragments present. Assuming that for a given magnetogram $N$ PIL fragments are detected, then the parameter $L_{tot}$ is the sum of all $N$ PIL fragment lengths $L_{j}$ detected at that instance, namely:

\begin{equation}
    L_{tot} = \sum_{j=1}^{N} L_{j}
\end{equation}
\phantom{...}

Once $L_{tot}$ is computed at all heights and timestamps for a given AR, we visualise the (i) PIL detection plots and (ii) stack-plots of $L_{tot}$ evolution at different heights above the photosphere for further analysis (see examples in Section \ref{ssec:vis}, for results of other ARs visit the Zenodo Project repository\footnote{{\url{https://zenodo.org/records/22738623}}} \cite{Biswal2026}). A brief discussion on uncertainties applicable to $L_{tot}$ is provided in Appendix \ref{app:Monte_Carlo}.


\clearpage
\newpage
\begin{figure}[h]
    \centering
    \includegraphics[width=1.0\linewidth]{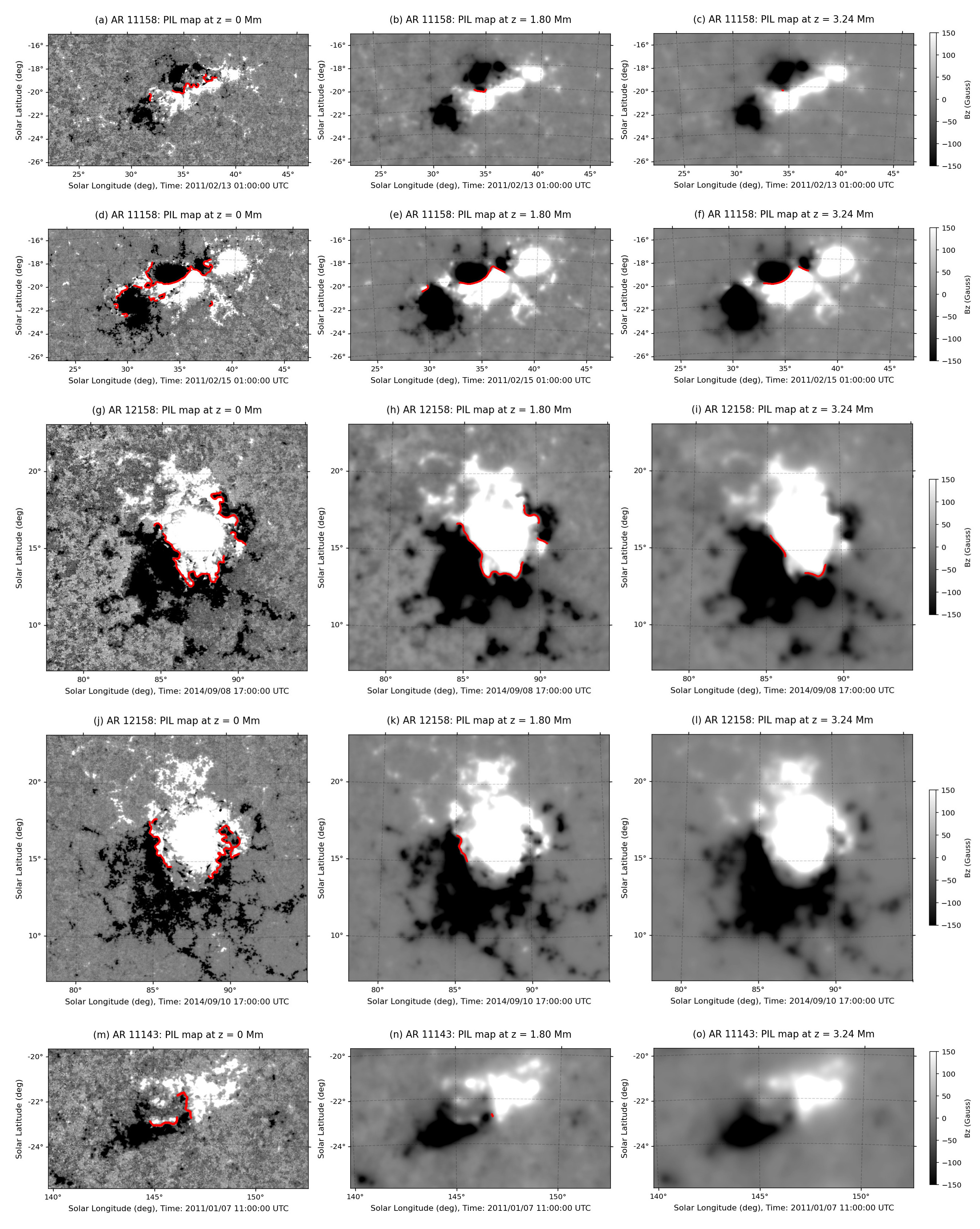}
    \caption{Plots showing detected PILs overlaid in red on the extrapolated magnetic field for SFARs - 11158 (a-f), 12158 (g-l) and NFAR 11143 (m-o) at a height of 0 Mm (col 1), 1.80 Mm (col 2) and 3.24 Mm (col 3); the colorbar thresholds are $\pm$ 150 G}
    \label{fig:XM-PILs}
    
\end{figure}

\clearpage
\newpage
\section{RESULTS}
\label{sec:rd}
\vspace{-0.4cm}
\phantom{...}
\subsection{Visualising PIL detection in the extrapolated heights}
\label{ssec:vis}

\phantom{...}

\begin{figure}[h]
\centering
\begin{subfigure}
  \centering
  \includegraphics[width=.325\linewidth]{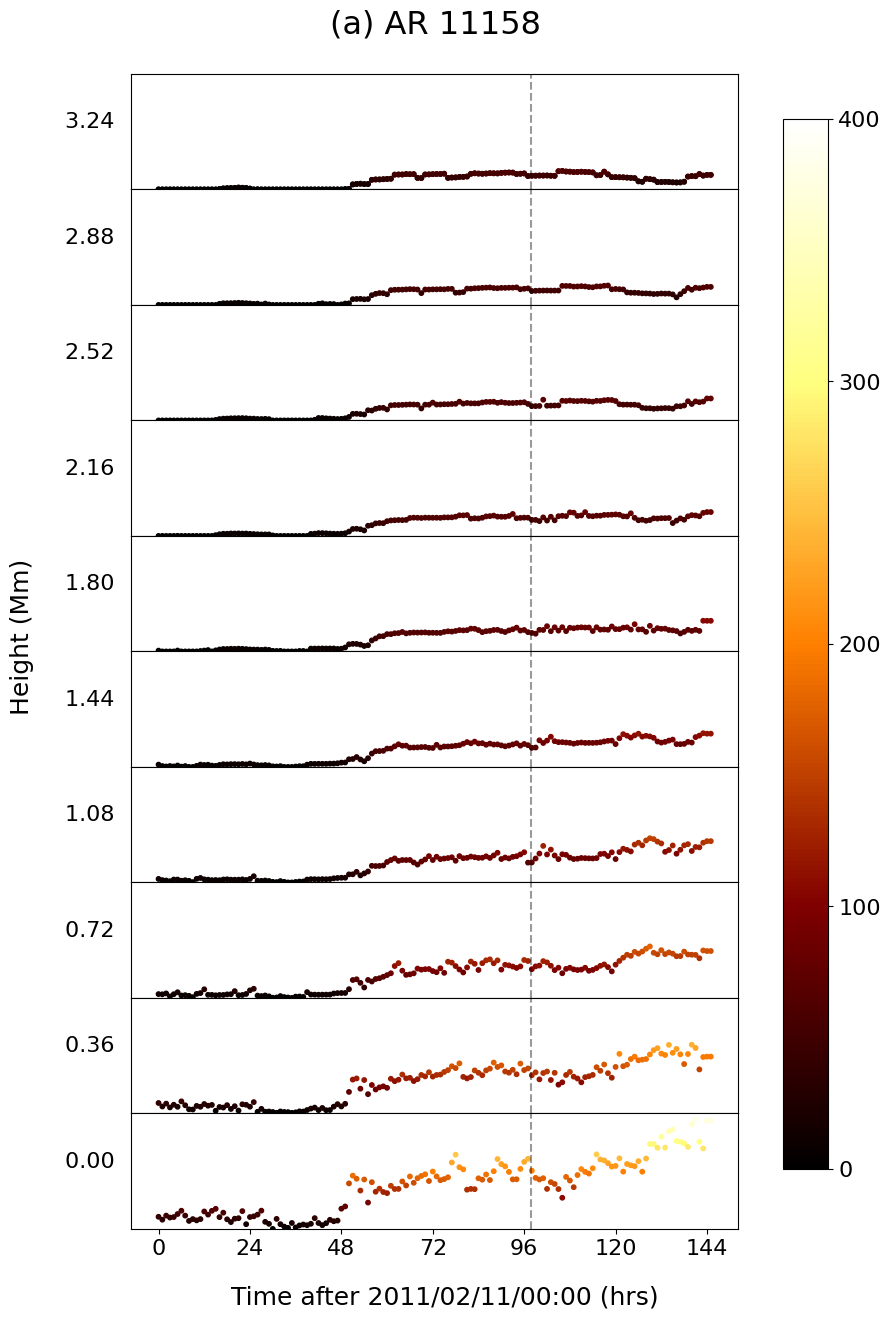}  
\end{subfigure}
\begin{subfigure}
  \centering
  \includegraphics[width=.325\linewidth]{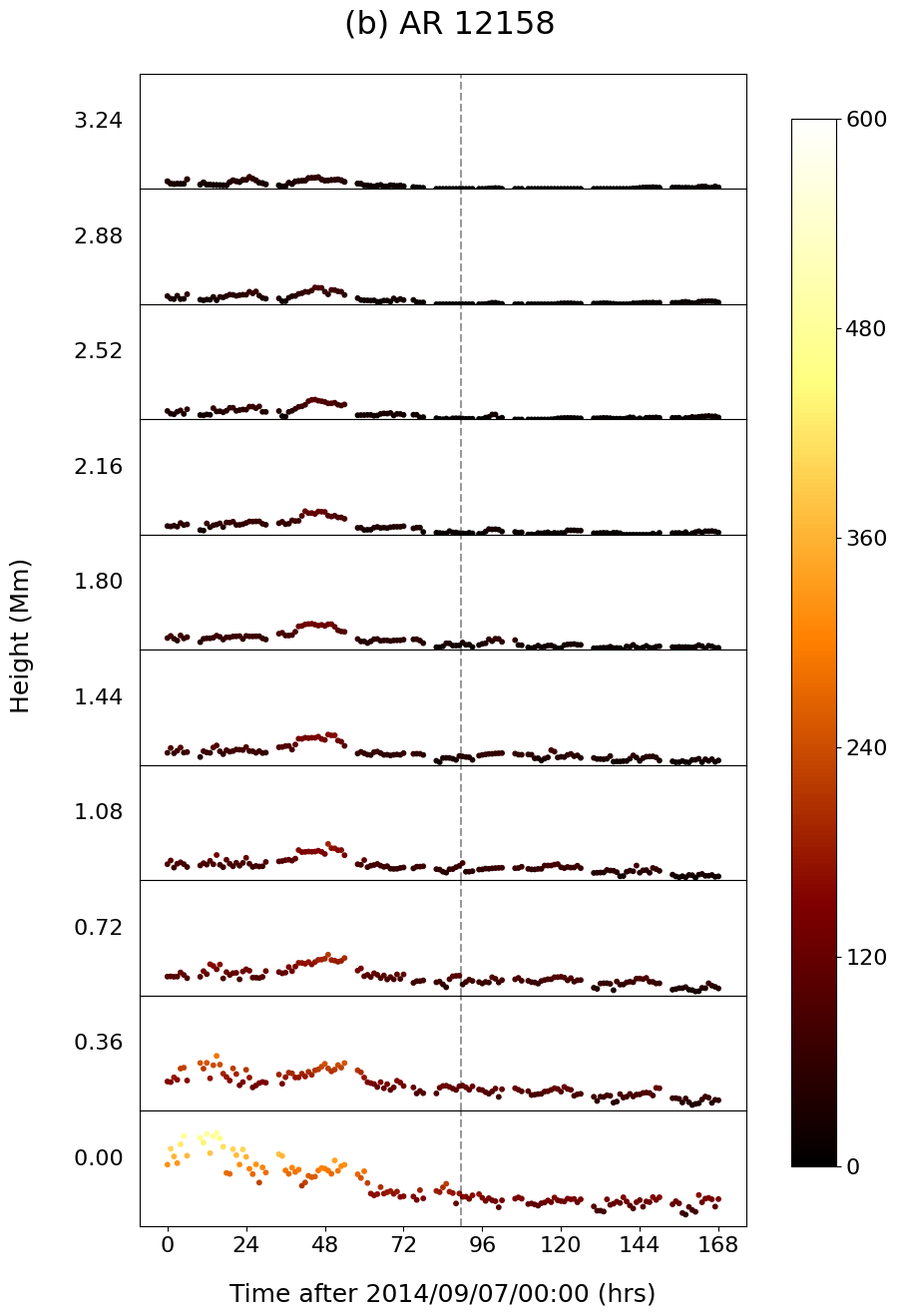}
\end{subfigure}
\begin{subfigure}
  \centering
  \includegraphics[width=.31\linewidth]{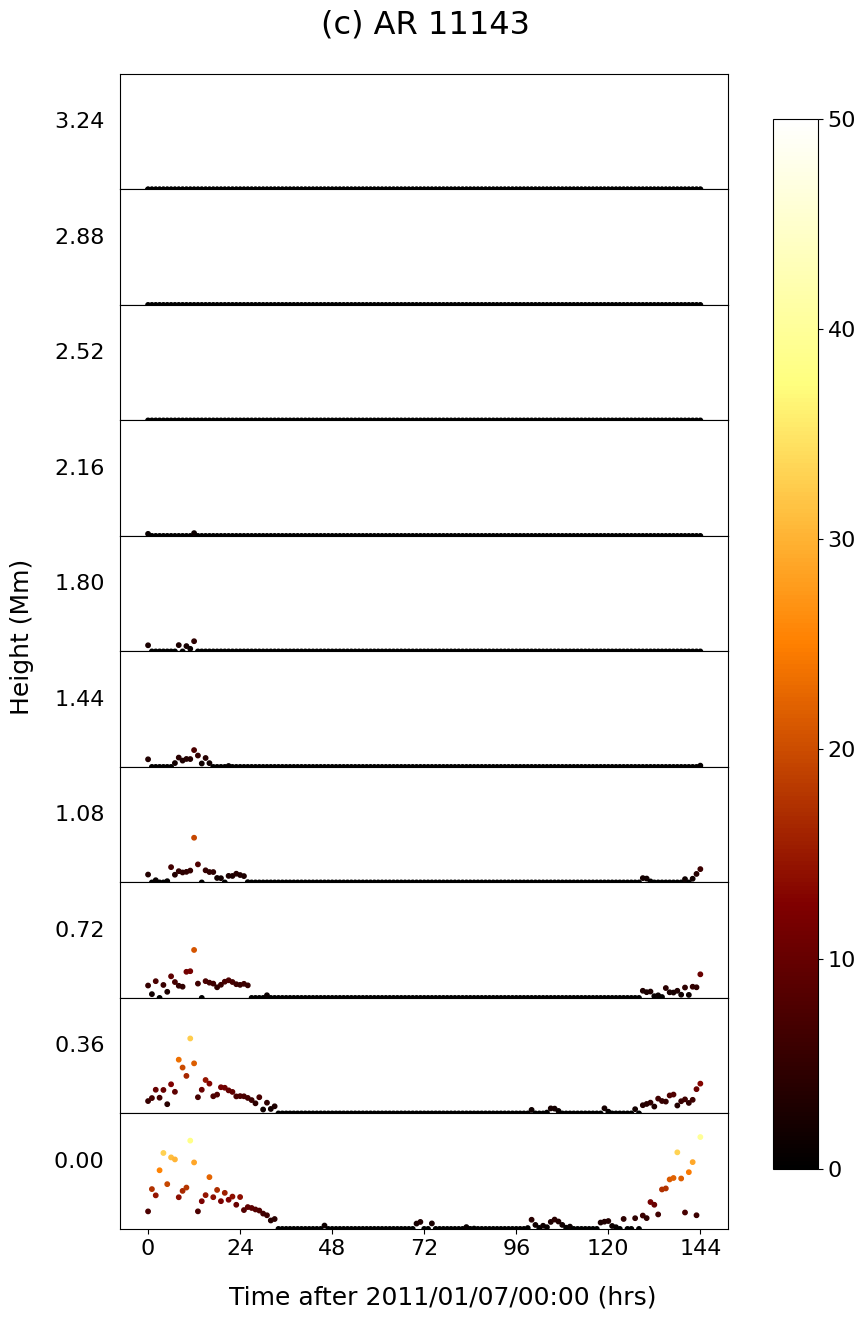}
\end{subfigure}

\caption{Multiple-height stack plots of $L_{tot}$ for (0-3.24 Mm) for (a) AR 11158 (b) AR 12158 (c) AR 11143 as a function of time; the colorbar in each case represents the magnitude of the $L_{tot}$ (in Mm); the colorbar's upper limit is set to 400 Mm for (a), 600 Mm for (b) and 50 Mm for (c); the dashed line in (a) and (b) represents the time of occurrence of the X-class flare}
\label{fig:Ltotstack}
\end{figure}

In this Section, we present examples of PIL detections (see Figure \ref{fig:XM-PILs}) using two SFARs (ARs 11158, 12158) and one NFAR (AR 11143). While Figure \ref{fig:XM-PILs} shows the PIL snapshots at specific points in time, Figure \ref{fig:Ltotstack} shows the time series data of $L_{tot}$ at different altitudes, computed from PF extrapolation, corresponding to the ARs visualised in Figure \ref{fig:XM-PILs}. We observe that the evolution of $L_{tot}$ with height is different for SFARs and NFARs; and no PILs were found above 2.52 Mm for the AR 11143 at any time (see Figure \ref{fig:Ltotstack}c), e.g. Figure \ref{fig:XM-PILs}o shows that the opposite polarities in AR 11143 have a gap between them. The PIL is not detected here as it does not exceed the control parameter thresholds. \\

Further, the two SFARs also exhibited different $L_{tot}$ evolution patterns. AR 11158, which is known to have exhibited (i) flux emergence (\citealp{Sun_2012}; \citealp{Vemareddy2015}; \citealp{Biswal2024}) and in which, (ii) the regions of opposite polarities moved towards each other during the interval of study, showed an overall increase in $L_{tot}$ (see Figures \ref{fig:Ltotstack}a \& \ref{fig:XM-PILs}a-f). Both these factors may have contributed to PIL growth, which can be observed at all heights (for e.g. compare Figures \ref{fig:XM-PILs}b \& \ref{fig:XM-PILs}e to check the growth at a height of 1.80 Mm). If we consider Figure \ref{fig:XM-PILs}c, about 48 hr prior to the X2.2 flare, we find that regions of opposite polarities barely touch each other ($L_{tot}$ $\sim$1 Mm). However on 15 February 2011 at 01:00 UTC, less than 1 hr prior to the flare, it can be be seen that these regions have converged and thus account for an increase in $L_{tot}$ to $\sim$46 Mm at a height of 3.24 Mm (see Figure \ref{fig:XM-PILs}f). Contrary to AR 11158, AR 12158, which is known to have shown (i) a decrease in unsigned flux in the interval of study \citep{Biswal2024}, and in which, (ii) the regions of opposite polarities moved away from each other during the interval of study, showed an overall decrease in $L_{tot}$ (see Figures \ref{fig:Ltotstack}b \& \ref{fig:XM-PILs}g-l). \\

\vspace{-0.4cm}

PF extrapolation smooths the magnetic field as one moves up in height (see Figure \ref{fig:XM-PILs}). This may have some advantages and disadvantages. The advantages are that (i) it makes the data less noisy and that (ii) it makes it easier to trace the movement of the regions having opposite polarities, with time. However, on the downside, smoothing may suppress smaller isolated fragments of PILs see, e.g., how some PIL fragments in Figure~\ref{fig:XM-PILs}d are missing in Figure~\ref{fig:XM-PILs}f). 

\newpage

\subsection{Decrease of $L_{tot}^*$ with height}
\label{ssec:attenuation}

\phantom{...}

\begin{figure}[h]
    \centering
    \includegraphics[width=0.95\linewidth]{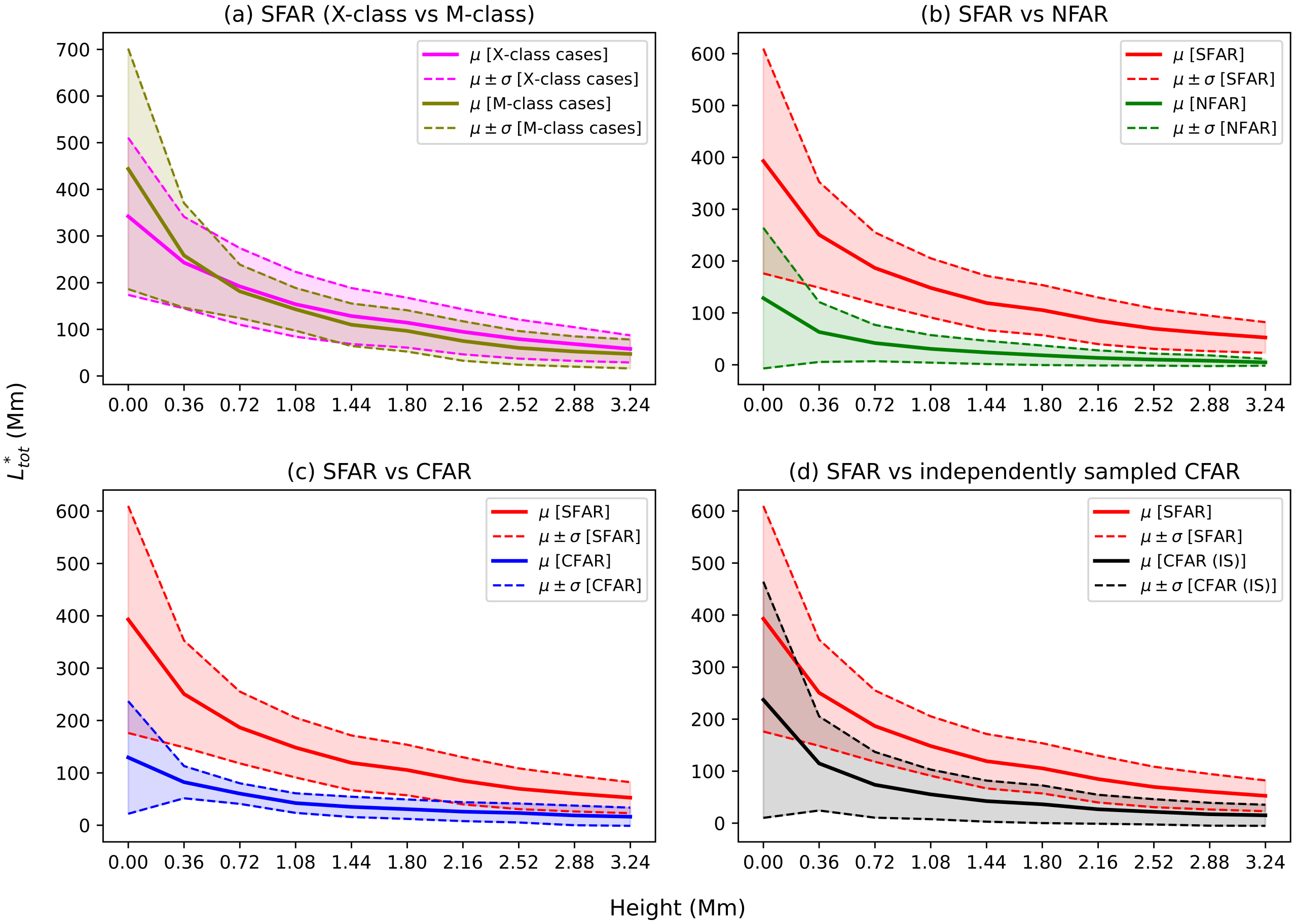}
    \caption{Profile of $L_{tot}^*$ decrease for various AR categories; the exact category-wise mean ($\mu$) and standard deviation ($\sigma$) values of $L_{tot}^*$ at each height can be found in Table \ref{table:statistics}}
    \label{fig:decay}
\end{figure}

Figure \ref{fig:decay}a shows the decrease of $L_{tot}^*$ with height above the photosphere for X-class flaring ARs when compared with M-class flaring ARs. Based on their mean ($\mu$) and standard deviation ($\sigma$) values, we find that these two groups cannot be easily distinguished. Hence for further analysis, we consider these two flaring groups as a single category of SFARs. Further Figures \ref{fig:decay}b, \ref{fig:decay}c and \ref{fig:decay}d show the variation of $L_{tot}^*$ with height for the NFARs, CFARs and independently sampled CFARs, respectively, when compared to SFARs. We find that a clear distinction exists between SFARs and NFARs at heights $\ge$ 0.36 Mm (see Figure \ref{fig:decay}b) and between SFARs and CFARs within a height range of 0.36-1.80 Mm (see Figure \ref{fig:decay}c), at least at $1 \sigma$ significance. These height ranges could serve as an OHR (optimal height range) for distinguishing SFARs from NFARs or CFARs, thereby demonstrating the importance of studying PILs in the PF-extrapolated configuration of ARs above the photosphere. \\

Further, the distribution of $L_{tot}^*$ at all heights for CFARs in terms of ($\mu,\sigma$) is not very different from their counterpart for NFARs, which are sampled independently by definition. This suggests that CFARs may have $L_{tot}^*$ values similar to that of NFARs, in the 48 hr period immediately preceding the first C-class flare. However, once these ARs evolve and progressively become more complex after the first C-class flare, they may have a higher chance of having $L_{tot}^*$ values similar to that of SFARs. This is evident from the partial overlap between the shaded regions in Figure \ref{fig:decay}d. Despite the partial overlap, $L_{tot}^*$ of SFARs, on average, is considerably greater (by a factor between 1.7 - 3.5, see Table \ref{table:statistics} for $\mu$ values at each height) than the independently sampled CFARs in general. If we make the same comparison between SFARs and NFARs, $L_{tot}^*$ differs by a factor between 3 - 11, depending upon height.  

\clearpage
\enlargethispage{3cm}
\vspace{-0.2cm}
\begin{table}[!ht]
\hskip-2.7cm
\begin{tabular}{| c | c | c | c | c | c | c | c | c | c | c |}
 \hline
 \multicolumn{11}{|c|}{} \\[-0.9em]
 \multicolumn{11}{|c|}{Mean values - $\mu (L_{tot}^*)$ (in Mm) as a function of height, listed columnwise} \\[0.4em]
 \hline
 \phantom{x} & \phantom{x} & \phantom{x} & \phantom{x} & \phantom{x} & \phantom{x} & \phantom{x} & \phantom{x} & \phantom{x}  & \phantom{x} & \phantom{x} \\ [-0.9em]
 AR Type & 0.00 Mm & 0.36 Mm & 0.72 Mm & 1.08 Mm & 1.44 Mm & 1.80 Mm & 2.16 Mm & 2.52 Mm & 2.88 Mm & 3.24 Mm \\[0.3em]
 \hline
 \phantom{x} & \phantom{x} & \phantom{x} & \phantom{x} & \phantom{x} & \phantom{x} & \phantom{x} & \phantom{x} & \phantom{x}  & \phantom{x} & \phantom{x} \\ [-1.2em]
 SFAR (X) & 341.99 & 242.92 & 191.87 & 153.60 & 128.35 & 114.16 & 94.52 & 78.91 & 68.27 & 57.90 \\
 SFAR (M) & 443.77 & 258.32 & 181.28 & 142.76 & 109.68 & 96.50 & 74.86 & 60.13 & 52.29 & 47.04 \\
 SFAR & 392.88 & 250.62 & 186.57 & 148.18 & 119.02 & 105.33 & 84.69 & 69.52 & 60.28 & 52.47 \\ 
 CFAR & 129.24 & 82.13 & 60.31 & 42.15 & 34.93 & 30.53 & 25.76 & 23.23 & 18.59 & 16.11 \\
 CFAR (IS) & 237.03 & 114.72 & 73.62 & 55.22 & 42.24 & 36.20 & 26.59 & 21.67 & 17.04 & 15.03 \\
 NFAR & 128.38 & 63.09 & 41.74 & 30.58 & 23.68 & 18.15 & 13.13 & 9.85 & 7.64 & 4.6 \\
 [0.3em]
 \hline
 \multicolumn{11}{|c|}{} \\[-0.9em]
 \multicolumn{11}{|c|}{Standard deviation values - $\sigma (L_{tot}^*)$ (in Mm) as a function of height, listed columnwise} \\[0.4em]
 \hline
 \phantom{x} & \phantom{x} & \phantom{x} & \phantom{x} & \phantom{x} & \phantom{x} & \phantom{x} & \phantom{x} & \phantom{x}  & \phantom{x} & \phantom{x} \\ [-0.9em]
 AR Type & 0.00 Mm & 0.36 Mm & 0.72 Mm & 1.08 Mm & 1.44 Mm & 1.80 Mm & 2.16 Mm & 2.52 Mm & 2.88 Mm & 3.24 Mm \\[0.3em]
 \hline
 \phantom{x} & \phantom{x} & \phantom{x} & \phantom{x} & \phantom{x} & \phantom{x} & \phantom{x} & \phantom{x} & \phantom{x}  & \phantom{x} & \phantom{x} \\ [-1.2em]
 SFAR (X) & 168.38 & 98.17 & 82.13 & 69.50 & 60.03 & 53.51 & 48.31 & 41.80 & 36.13 & 29.12 \\
 SFAR (M) & 257.74 & 112.06 & 57.26 & 45.73 & 45.56 & 44.29 & 42.30 & 36.07 & 32.48 & 31.03 \\ 
 SFAR & 216.78 & 102.08 & 68.61 & 57.11 & 52.37 & 48.32 & 45.02 & 38.94 & 34.19 & 29.61 \\
 CFAR & 107.53 & 30.78 & 19.66 & 18.72 & 19.42 & 18.59 & 18.05 & 18.12 & 18.74 & 17.40 \\
 CFAR (IS) & 226.98 & 90.50 & 63.19 & 47.65 & 39.49 & 36.20 & 27.82 & 24.33 & 21.96 & 20.27 \\
 NFAR & 135.59 & 57.78 & 35.04 & 26.55 & 22.42 & 18.82 & 14.6 & 11.6 & 10.36 & 6.51 \\ [0.3em]
 \hline
 \end{tabular}
\caption{Category-wise $\mu$ and $\sigma$ values of $L_{tot}^*$ at each height; SFAR (X) represents X-class flaring ARs; SFAR (M) represents M-class flaring ARs; CFAR (IS) represents independently sampled CFARs}
\label{table:statistics}
\end{table}
\vspace{-0.8cm}
\begin{figure}[!ht]
    \centering \includegraphics[width=1.0\linewidth]{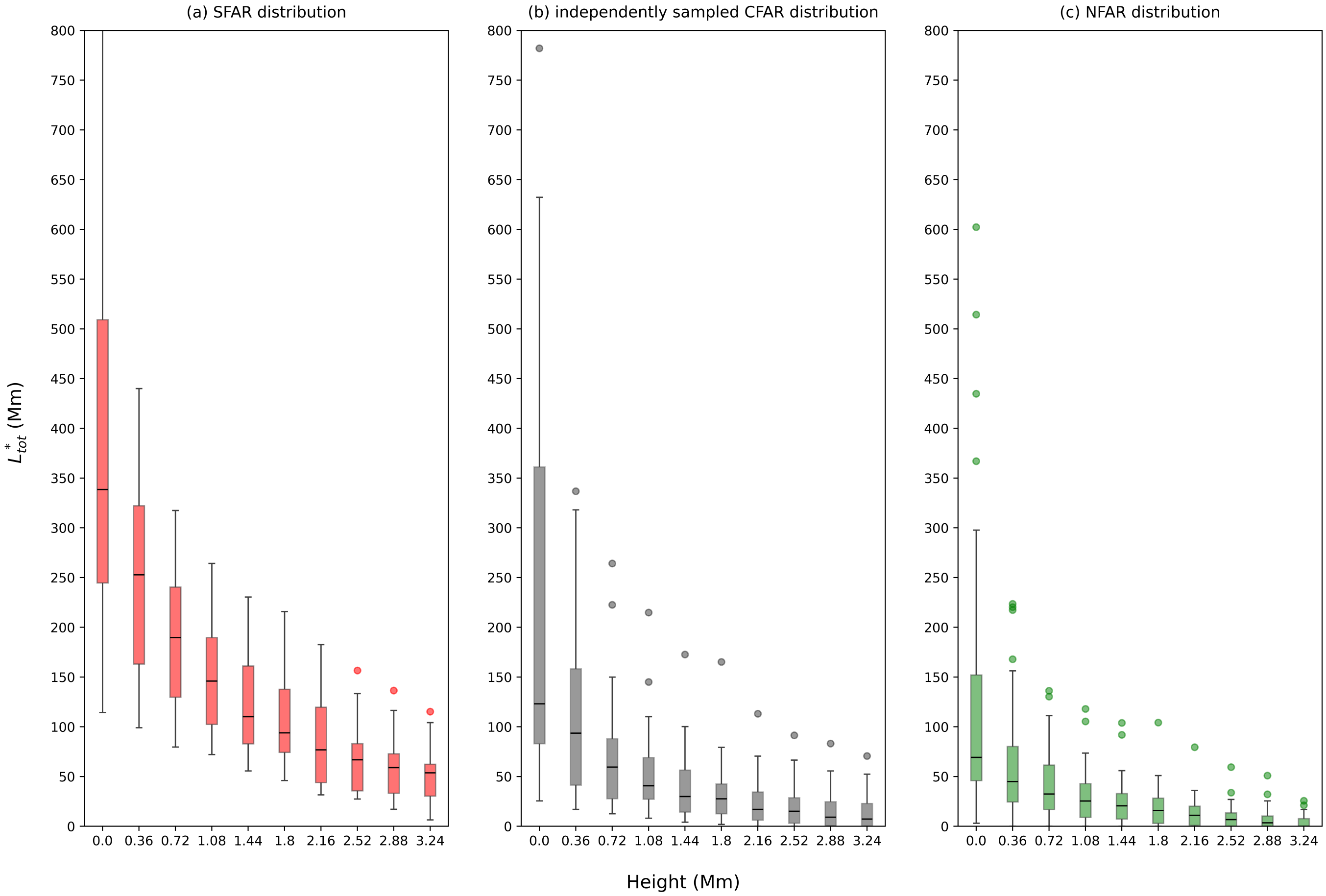}
    \caption{{Boxplots of $L_{tot}^*$ as a function of height for (a) SFARs, (b) independently sampled CFARs and (c) NFARs; the lower and upper boundaries of each box denote the 25th ($Q_1$) and 75th ($Q_3$) percentiles; the median is shown by a black line inside each box; the whiskers denote $L_{tot}^* =$  1.5 × IQR (inter-quartile range = $Q_3-Q_1$) from $Q_1$ and $Q_3$; the coloured dots are outliers}}
    \label{fig:boxplot}
\end{figure}


\clearpage
\newpage

\subsection{Optimal height for distinguishing SFARs from NFARs and independently sampled CFARs}
\label{ssec:samp}

\phantom{...}

Heights $\ge$ 0.36 Mm may serve as an OHR to distinguish SFARs from NFARs, as mentioned previously in Section~\ref{ssec:attenuation}. Similarly, a height range of 0.36-1.80 Mm may help distinguish between SFARs from CFARs, which are a subset of the independently sampled CFARs. Since we would also like to investigate how well we could separate between independently sampled CFARs and SFARs even though their $L_{tot}^*$-decrease profiles show a partial overlap, we made a box-plot analysis of the distribution as a function of altitude (Figure \ref{fig:boxplot}). The box-plot shows a clear separation between the IQRs (inter-quartile ranges) of SFARs and independently sampled CFARs at all heights $\ge$ 0.36 Mm. The IQRs here correspond to the middle 50\% of the data and is denoted by the span of the coloured boxes in Figure \ref{fig:boxplot}. This finding once again, demonstrates the importance of studying PILs in the extrapolated topology of ARs above the photosphere. Similarly, we find that the IQRs of SFARs and NFARs are well separated at all heights in our study. \\

The boxplots in Figure~\ref{fig:boxplot} also indicate that the heights 1.80 Mm and 2.16 Mm contain the least number of outliers if we consider all the three groups - SFARs, NFARs and independently sampled CFARs, together. Thus these heights could be most suitable within the OHR for distinguishing SFARs from NFARs and independently sampled CFARs. To verify this, we take the minimum $L_{tot}^*$ value for SFARs at all heights as a threshold ($L_{th}$) and check how many cases within 45 NFAR samples and 23 independently sampled CFARs samples did not exceed this threshold (see Table \ref{table:successrate}). Our findings reveal that the 1.80 Mm height is the most optimal height (within the OHR) to separate out SFARs from independently sampled CFARs and NFARs. We find that $\sim$96\% of NFARs and $\sim$78\% of independently sampled CFARs are less than $L_{th}$ $\sim$46 Mm at a height of 1.80 Mm above the photosphere. \\

\begin{table}[h!]
\hskip-2.95cm
\begin{tabular}{| c | c | c | c | c | c | c | c | c | c | c |}
 \hline
 \multicolumn{11}{|c|}{} \\[-1.4em]
 \phantom{x} & \phantom{x} & \phantom{x} & \phantom{x} & \phantom{x} & \phantom{x} & \phantom{x} & \phantom{x} & \phantom{x}  & \phantom{x} & \phantom{x} \\ [-1.2em]
 Height (Mm) & 0.00 & 0.36 & 0.72 & 1.08 & 1.44 & \textbf{1.80} & 2.16 & 2.52 & 2.88 & 3.24 \\[0.3em]
 \hline
 \phantom{x} & \phantom{x} & \phantom{x} & \phantom{x} & \phantom{x} & \phantom{x} & \phantom{x} & \phantom{x} & \phantom{x}  & \phantom{x} & \phantom{x} \\ [-1.2em]
 $L_{th}$ (Mm) & 114.24 & 99.03 & 79.68 & 72.07 & 55.67 & \textbf{45.99} & 31.62 & 27.38 & 17.12 & 6.45 \\ [0.3em]
 \hline
 \phantom{x} & \phantom{x} & \phantom{x} & \phantom{x} & \phantom{x} & \phantom{x} & \phantom{x} & \phantom{x} & \phantom{x}  & \phantom{x} & \phantom{x} \\ [-1.2em]
 NFAR & 28/45 & 37/45 & 38/45 & 42/45 & 42/45 & \textbf{43/45} & 43/45 & 43/45 & 39/45 & 31/45 \\
  $(L_{tot}^* < L_{th})\%$ & 62.22\% & 82.22\% & 84.44\% & 93.33\% & 93.33\% & \textbf{95.56\%} & 95.56\% & 95.56\% & 86.67\% & 68.89\% \\ [0.3em]
 \hline
 \phantom{x} & \phantom{x} & \phantom{x} & \phantom{x} & \phantom{x} & \phantom{x} & \phantom{x} & \phantom{x} & \phantom{x}  & \phantom{x} & \phantom{x} \\ [-1.2em]
 CFAR (IS) & 11/23 & 12/23 & 16/23 & 17/23 & 17/23 & \textbf{18/23} & 16/23 & 16/23 & 15/23 & 11/23 \\ 
 $(L_{tot}^* < L_{th})\%$ & 47.83\% & 52.17\% & 69.57\% & 73.91\% & 73.91\% & \textbf{78.26\%} & 69.57\% & 69.57\% & 65.22\% & 47.83\% \\
 [0.3em]
 \hline
 \end{tabular}
\vskip0.5cm
\caption{Variation of $L_{th}$ i.e. the minimum $L_{tot}^*$ value (row 2) observed for SFARs with height (row 1) and details of the percentage of NFARs and independently sampled CFARs which satisfied the condition: $L_{tot}^* < L_{th}$ at each height (rows 3 and 4, respectively). The best distinction is achieved at a height of 1.80 Mm (column in bold).}
\label{table:successrate}
\end{table}

\vspace{-0.6cm}
\subsection{Optimal height for identifying population clusters from the current data sample}
\label{ssec:pop}

\phantom{...}

In the boxplots of Figure~\ref{fig:boxplot}, it is observed that the upper whisker is in general longer than the lower whisker for all three groups of ARs at all heights. This is an indicator of the low-value skewness in the distributions and suggests that our data is not normally distributed. In this context, we present a method to identify and distinguish between different population clusters (for e.g. SFARs and NFARs) using the one-sided Chebyshev's inequality \citep{Lin2011} applying to all kinds of distributions. The inequality states that if $X$ is a random variable with mean $\mu$ and standard deviation $\sigma$, then for $\lambda$ $>$ 0 the following are true ($P$ denoted probability):

\begin{equation}
    P(X \le \mu + \lambda) \ge \frac{\lambda^2}{\lambda^2 + \sigma^2} \le P(X \ge \mu - \lambda)
    \label{eq:can1}
\end{equation}

\phantom{...}

If we analyse Figure~\ref{fig:decay}a, we find that the $\mu+\sigma$ line for NFARs is well below the $\mu-\sigma$ line for SFARs. It is intuitive to note here that if we multiply $\sigma$ by a fixed factor $k > 1$, then the gap between the two shaded regions can be minimised. This gap can be eliminated if:

\begin{equation}
    k = \frac{\mu(SFAR) - \mu(NFAR)}{\sigma(SFAR) + \sigma(NFAR)}
    \label{eq:k}
\end{equation}

\phantom{...}

Using Equation \ref{eq:k}, we find $k = \{1.17, 1.40, 1.41, 1.27, 1.30, 1.20, 1.18, 1.18, 1.33\}$ for the nine heights between 0.36 Mm and 3.24 Mm, in ascending order. The maximum value of $k = 1.41$ is obtained at 1.08 Mm height. At this height, the $\mu+1.41\sigma$ level for NFARs and the $\mu-1.41\sigma$ level for SFARs are both $\sim$68 Mm. Letting $\lambda = 1.41 \sigma$ in Equation \ref{eq:can1} in each case, we have:

\begin{equation}
    P(0 \le L_{tot}^*(NFAR) \le 68) \ge 66\%
    \label{eq:can3}
\end{equation}

\begin{equation}
    P(L_{tot}^*(SFAR) \ge 68) \ge 66\%
    \label{eq:can4}
\end{equation}

\phantom{...}

This implies that if our data samples for SFARs and NFARs were to be considered as fairly good representatives for the actual populations of SFARs and NFARs, then at a height of 1.08 Mm above the photosphere, a threshold of $L_{tot}^*$ = 68 Mm may provide classification-type probabilistic guarantees that at least 66\% of the SFAR population must be above this threshold and at least 66\% of the NFAR population must be below it. \\

\section{DISCUSSION AND CONCLUSIONS}
\label{sec:conc}

\phantom{...}
\vspace{-0.3cm}
\subsection{Implications on flare predictability and forecasting}\label{ssec:imp}

\phantom{...}

Even though it is known that PILs are intimately connected to solar eruptive phenomena and PIL length has been correlated to flare and eruptive flare (i.e., flare-associated CME) likelihoods (\citealp{Falconer2002}; \citealp{Falconer2003}; \citealp{Kontogiannis2019}), actual PIL lengths, especially above the photosphere, have been barely discussed. As per \cite{Falconer2002}, the PIL parameter $L_{SS}$ (called the strong shear length), which is determined by calculating the length of the segments on the main neutral line (i.e. the longest PIL) where the horizontal magnetic field $B_h$ $>$ 150 G and the shear angle $\theta$ $>$ 45$^\circ$, is well correlated to CME productivity within a $\pm$ 48 hr time window. \cite{Falconer2003} have shown that $L_{SS}$ is well correlated to another PIL parameter $L_{SG}$ (called the strong gradient length) with a confidence level exceeding 95\%, which is determined by calculating the length of the segments on the main neutral line where $B_h$ $>$ 150 G and the absolute value of the horizontal gradient of the line of sight magnetic field $\nabla_hB_{los}$ $>$ 50 G/Mm. \cite{Falconer2003} have also shown that if $L_{SS}$ exceeds 62 Mm or if $L_{SG}$ exceeds 36 Mm, then CMEs are to be expected (accuracy score $\sim$88\%). \\

Taking these previous results into consideration, we can say that the PIL parameter of focus in this paper - $L_{tot}$ is slightly different from $L_{SG}$ as it includes all PIL segments in an AR, not just the central PIL. While $L_{SG}$ takes into account the vertical magnetic field in the form of $\nabla_hB_{los}$, $L_{tot}$ does so through $B_{zg}$. Even though $L_{tot}$ is computed using less stricter threshold in $B_h$ compared to $L_{SG}$, this difference may be justified by the fact that our approach computes PILs not just at the photosphere but also at heights above it. As we move up in height, the magnetic field values also become weaker, thereby plausibly requiring weaker thresholds for any meaningful PIL detection. Overall, however, the major similarity between $L_{tot}$ and $L_{SG}$ is that both these parameters use the vertical and horizontal components of the magnetic field as physics-based control parameters for PIL detection. In this context, when detecting PILs using a uniform set of thresholds at each height within a height range of 0-3.24 Mm above the photosphere, $L_{tot}$ may be considered as a useful alternative to $L_{SG}$. We find that $L_{tot}^*$, i.e. the maximum value of $L_{tot}$ over a 48 hr interval immediately before the first X/M-class flare is in general longer than the respective for ARs that do not host any X/M-class flares. \\ 

Using descriptive statistics, we find that a height of 1.8 Mm above the photosphere may be the most suitable height to distinguishing SFARs from NFARs and independently sampled CFARs. At this height, $\sim$78\% of independnently sampled CFARs and $\sim$96\% of NFARs, $L_{tot}^*$ was less than 46 Mm. While these results appear promising at this stage to develop a binary (YES/NO) model to detect SFARs at an optimal height of 1.8 Mm, we acknowledge the caveat of a limited sample of ARs. Working on a larger AR sample will help refine our existing results. This is reserved for a future work on this topic. A large scale study could provide information on (i) the relative natural occurrence frequencies of SFARs ($f_1$), CFARs ($f_2$) and NFARs ($f_3$) and (ii) the probabilities of $L_{tot}^* \ge L_{th}$ for SFARs ($p_1$), CFARs ($p_2$) and NFARs ($p_3$), where $L_{th}$ is a threshold based on some relevant statistical metric (for e.g. minimum value, 5\%ile value etc.) pertaining to the $L_{tot}^*$ distribution for SFARs at a given optimal height. Based on this information, the overall problem of forecasting M/X-class flares over a 48 hr window may be translated into a simpler problem of Bayesian inference. In real-time, if we find $L_{tot}^* \ge L_{th}$ at the optimal height, then the probability that it would produce an M/X-class flare within 48 hr shall be: 

\begin{equation}
    p = \frac{f_1p_1}{f_1p_1 + f_2p_2 + f_3p_3}
\end{equation}

\phantom{...}

Using the one-sided Chebyshev inequality, we inferred that a height of 1.08 Mm above the photosphere may be the most suitable height to demarcate the population of SFARs from NFARs. At this height, at least 66\% of SFARs have $L_{tot}^* \ge$ 68 Mm and at least 66\% of NFARs have $L_{tot}^* \le$ 68 Mm. In this regard, a more rigorous statistical study could be helpful for a more accurate statistical separation of the different populations. \\

\subsection{Summary}\label{ssec:summary}

\phantom{...}

Some general conclusions of this study are summarised below:

\begin{itemize}

    \item[$\bullet$\hspace{0.3cm}] While examining the evolution of precursors or predictors such as the $W_{GM}$ morphological parameter \citep{Korsos2020}, magnetic helicity \citep{Korsos2022} and $R$-value \citep{Biswal2024} in the lower solar atmosphere are useful for maximising the flare prediction lead time, analysing the magnitudes of $L_{tot}$ and $L_{tot}^*$ in the extrapolated topology of ARs within an OHR in the lower solar atmosphere may offer critical complementary information concerning the prediction of strong flares.

    \item[$\bullet$\hspace{0.3cm}] {Specifically, $L_{tot}$ - computed with an uniform set of thresholds (same as those originally intended for the photosphere by \cite{Georgoulis2021} in the FLARECAST project) - appears to be a useful parameter to study PILs in the LSA (up to about 3 Mm above the photosphere).} The overall approach of analysing $L_{tot}$ as a function of height may be used (i) as a method to distinguish between SFARs and NFARs; (ii) as a technique to determine suitable thresholds of $L_{tot}^*$ as a function of height, which may serve as a basis for demarcating the population clusters of SFARs, NFARs and independently sampled CFARs; and (iii) potentially to estimate the probability of occurrence of M/X-class flares within a 48 hr window, provided a statistical study can establish the natural occurrence frequencies of SFARs, NFARs and independently sampled CFARs as well as the likelihood of each category exhibiting $L_{tot}^*$ values above the thresholds at each height. 

    \item[$\bullet$\hspace{0.3cm}] The $L_{tot}^*$ decrease profiles show a clearer separation between SFARs and NFARs at heights $\ge$ 0.36 Mm as compared to the photosphere, thereby highlighting the importance of studying the extrapolated magnetic configuration of ARs. In this regard, it was also seen that for CFARs, the 48 hr period immediately before the first flare resembled NFARs in terms of $L_{tot}^*$, and was quite different from SFARs between a height range of 0.36-1.80 Mm. However, the $L_{tot}^*$ decrease profile of independently sampled CFARs showed a partial overlap with SFARs at all heights, and exhibited an overall pattern that is intermediate to SFARs and NFARs. 

    \item[$\bullet$\hspace{0.3cm}] We used two different approaches to find the most optimal height (within the OHR) which may bring about the best possible distinction between SFARs and NFARs. While the approach centred around descriptive statistics of our data sample shows that the best height is about 1.8 Mm above the photosphere, the other approach suggests that at about 1 Mm above the photosphere may be the most suitable to demarcate clusters of SFARs and NFARs at a population level.

    \item[$\bullet$\hspace{0.3cm}] It is promising to note from our dataset of 40 ARs and 92 cases (i.e. 48 hr time series) that $\sim$78\% of independently sampled CFARs and $\sim$96\% of NFARs had $L_{tot}^* <$ 46 Mm at 1.8 Mm height.    
    
\end{itemize}

We plan to extend our calculations, as a continuation of this study, to a much larger statistical study and other methods of extrapolations, which (i) may provide more accurate thresholds in the OHR and (ii) may help develop probabilistic models based on PIL lengths (i.e. $L_{tot}^*$) to forecast M/X-class flares. 

\clearpage
\newpage

\section{Acknowledgement}

\phantom{...}

This research work is supported by Space Weather Awareness Training Network (SWATNet) and it comes under the aegis of the  project: 'Three-dimensional solar flare prediction' (\href{https://swatnet.eu/}{https://swatnet.eu/}). SWATNet has received funding from the European Union’s Horizon 2020 research and innovation programme under the Marie Sklodowska-Curie Grant Agreement No. 955620. SB acknowledges the support received from the University of Sheffield Publication Scholarship Programme. MBK, RE and MG acknowledge support from ISSI-BJ ("Step forward in solar flare and coronal mass ejection (CME) predicting"). RE is grateful to STFC (UK, grant number ST/M000826/1) and PIFI (China, grant number No. 2024PVA0043). MBK is grateful for the Leverhulme Trust Found ECF-2023-271. MBK acknowledges support by UNKP-22-4-II-ELTE-186, ELTE Hungary. RE and MBK  also thank for the support received from NKFIH OTKA (Hungary, grant No. K142987). This work was also supported by the NKFIH Excellence Grant TKP2021-NKTA-64.  AN and SP acknowledge support by the ERC Synergy Grant ‘Whole Sun’ (GAN: 810218). \\

\newpage

\appendix

\section{Analysed Dataset} 
\label{app:dataset}

\phantom{...}

This Section contains the full list of all 92 different time series studied for X/M/C-class flaring ARs (see Table \ref{table:dataset1}), independently sampled CFARs (see Table \ref{table:dataset2}) and NFARs (see Table \ref{table:dataset2}) as referenced in Section \ref{sec:ardataset}. \\

{\begin{table}[h]
\hskip-1cm
\begin{tabular}{| c | c | c | c | c | c | c |}
 \hline
 \multicolumn{7}{|c|}{} \\
 [-0.9em]
 \multicolumn{7}{|c|}{Observed data for SFARs} \\[0.4 em]
 \hline
 \phantom{x} & \phantom{x} & \phantom{x} & \phantom{x} & \phantom{x} & \phantom{x} & \phantom{x} \\ [-0.9em]
No. & AR & SHARP No. & Interval $T_{start}$ & Interval $T_{end}$ & Flare Time & Flare Class \\
[0.3em]
\hline
 01 & 11158 & 377 & 2011/02/13 01:00 & 2011/02/15 01:00 & 2011/02/15 01:44 & X2.2 \\
 02 & 11166 & 401 & 2011/03/07 23:00 & 2011/03/09 23:00 & 2011/03/09 23:13 & X1.5 \\
 03 & 11283 & 833 & 2011/09/04 22:00 & 2011/09/06 22:00 & 2011/09/06 22:12 & X2.1 \\
 04 & 11520 & 1834 & 2012/07/10 15:00 & 2012/07/12 15:00 & 2012/07/12 15:37 & X1.4 \\
 05 & 12017 & 3894 & 2014/03/27 17:00 & 2014/03/29 17:00 & 2014/03/29 17:35 & X1.0 \\
 06 & 12158 & 4536 & 2014/09/08 17:00 & 2014/09/10 17:00 & 2014/09/10 17:21 & X1.6 \\
 07 & 12297 & 5298 & 2015/03/09 16:00 & 2015/03/11 16:00 & 2015/03/11 16:11 & X2.1 \\
 08 & 12673 & 7115 & 2017/09/04 09:00 & 2017/09/06 09:00 & 2017/09/06 08:57 & X2.2 \\
 [0.3em]
 \hline
 09 & 11620 & 2227 & 2012/11/26 21:00 & 2012/11/28 21:00 & 2012/11/28 21:20 & M2.2 \\
 10 & 11719 & 2635 & 2013/04/09 06:00 & 2013/04/11 06:00 & 2013/04/11 06:55 & M6.5 \\
 11 & 11731 & 2693 & 2013/04/30 04:00 & 2013/05/02 04:00 & 2013/05/02 04:58 & M1.1 \\
 12 & 11818 & 3056 & 2013/08/15 18:00 & 2013/08/17 18:00 & 2013/08/17 18:16 & M3.3 \\
 13 & 12253 & 5011 & 2015/01/01 09:00 & 2015/01/03 09:00 & 2015/01/03 09:40 & M1.1 \\
 14 & 12403 & 5885 & 2015/08/19 01:00 & 2015/08/21 01:00 & 2015/08/21 01:56 & M1.2 \\
 15 & 12422 & 5983 & 2015/09/25 20:00 & 2015/09/27 20:00 & 2015/09/27 20:54 & M1.2 \\
 16 & 12497 & 6327 & 2016/02/10 10:00 & 2016/02/12 10:00 & 2016/02/12 10:36 & M1.0 \\[0.3em]
 \hline
 \multicolumn{7}{|c|}{} \\
 [-0.9em]
 \multicolumn{7}{|c|}{Observed data for CFARs} \\[0.4 em]
 \hline
 17 & 11440 & 1484 & 2012/03/19 12:00 & 2012/03/21 12:00 & 2012/03/21 12:38 & C2.9 \\
 18 & 11553 & 1959 & 2012/08/27 23:00 & 2012/08/29 23:00 & 2012/08/29 23:40 & C1.1 \\
 19 & 11640 & 2337 & 2012/12/30 08:00 & 2013/01/01 08:00 & 2013/01/01 08:47 & C1.2 \\
 20 & 11730 & 2691 & 2013/04/27 20:00 & 2013/04/29 20:00 & 2013/04/29 20:22 & C5.9 \\
 21 & 12255 & 5022 & 2015/01/09 14:00 & 2015/01/11 14:00 & 2015/01/11 14:42 & C1.6 \\
 22 & 12293 & 5249 & 2015/03/02 13:00 & 2015/03/04 13:00 & 2015/03/04 13:35 & C1.1 \\
 23 & 12353 & 5596 & 2015/05/21 03:00 & 2015/05/21 03:00 & 2015/05/23 03:27 & C1.0 \\
 24 & 12552 & 6599 & 2016/06/07 12:00 & 2016/06/09 12:00 & 2016/06/09 12:28 & C1.4 \\[0.3em]
 \hline
\end{tabular}
\vspace{0.15cm}

\caption{Table listing the studied flaring ARs with assisting information such as NOAA number, SHARP number, observation time window ($T_{start}$ denoting the start time and $T_{end}$ denoting the end time), the time of flare occurrence and GOES flare class}
\label{table:dataset1}
\end{table}}


\begin{table}[h]
\begin{centering}

\begin{tabular}{| c | c | c | c | c || c | c | c | c | c |}
 \hline
 \multicolumn{10}{|c|}{} \\
 [-0.9em]
 \multicolumn{10}{|c|}{Observed data for independently sampled CFARs} \\[0.4 em]
 \hline
 \phantom{x} & \phantom{x} & \phantom{x} & \phantom{x} & \phantom{x}
 & \phantom{x} & \phantom{x} & \phantom{x} & \phantom{x} & \phantom{x} \\ [-0.9em]

 No. & AR & SHARP No. & Interval $T_{start}$ & Interval $T_{end}$
 & No. & AR & SHARP No. & Interval $T_{start}$ & Interval $T_{end}$ \\
 [0.3em]
 \hline

 1 & 11440-1 & 1484 & 2012/03/19 & 2012/03/21
 & 13 & 11730-3 & 2691 & 2013/04/30 & 2013/05/02 \\
 2 & 11440-2 & 1484 & 2012/03/21 & 2012/03/23
 & 14 & 12255-1 & 5022 & 2015/01/06 & 2015/01/08 \\
 3 & 11440-3 & 1484 & 2012/03/23 & 2012/03/25
 & 15 & 12255-2 & 5022 & 2015/01/08 & 2015/01/10 \\
 4 & 11553-1 & 1959 & 2012/08/24 & 2012/08/26
 & 16 & 12255-3 & 5022 & 2015/01/10 & 2015/01/12 \\
 5 & 11553-2 & 1959 & 2012/08/26 & 2012/08/28
 & 17 & 12293-1 & 5249 & 2015/02/27 & 2015/03/01 \\
 6 & 11553-3 & 1959 & 2012/08/28 & 2012/08/30
 & 18 & 12293-2 & 5249 & 2015/03/01 & 2015/03/03 \\
 7 & 11553-4 & 1959 & 2012/08/30 & 2012/09/01
 & 19 & 12293-3 & 5249 & 2015/03/03 & 2015/03/05 \\
 8 & 11640-1 & 2337 & 2012/12/30 & 2013/01/01
 & 20 & 12353-1 & 5596 & 2015/05/21 & 2015/05/23 \\
 9 & 11640-2 & 2337 & 2013/01/01 & 2013/01/03
 & 21 & 12353-2 & 5596 & 2015/05/23 & 2015/05/25 \\
 10 & 11640-3 & 2337 & 2013/01/03 & 2013/01/05
 & 22 & 12353-3 & 5596 & 2015/05/25 & 2015/05/27 \\
 11 & 11730-1 & 2691 & 2013/04/26 & 2013/04/28
 & 23 & 12552-1 & 6599 & 2016/06/08 & 2016/08/10 \\
 12 & 11730-2 & 2691 & 2013/04/28 & 2013/04/30
 & & & & & \\ [0.3em]

 \hline
  \multicolumn{10}{|c|}{} \\
 [-0.9em]
 \multicolumn{10}{|c|}{Observed Data for NFARs} \\[0.4 em]
 \hline
 \phantom{x} & \phantom{x} & \phantom{x} & \phantom{x} & \phantom{x} & \phantom{x} & \phantom{x} & \phantom{x} & \phantom{x} & \phantom{x} \\ [-0.9em]

 No. & AR & SHARP No. & Interval $T_{start}$ & Interval $T_{end}$
 & No. & AR & SHARP No. & Interval $T_{start}$ & Interval $T_{end}$ \\
 [0.3em]
 \hline

  1 & 11143-1 & 335 & 2011/01/07 & 2011/01/09
 & 24 & 11957-3 & 3648 & 2014/01/22 & 2014/01/24 \\
 2 & 11143-2 & 335 & 2011/01/09 & 2011/01/11
 & 25 & 11957-4 & 3648 & 2014/01/24 & 2014/01/26 \\
 3 & 11143-3 & 335 & 2011/01/11 & 2011/01/13
 & 26 & 12064-1 & 4135 & 2014/05/15 & 2014/05/17 \\
 4 & 11416-1 & 1389 & 2012/02/09 & 2012/02/11
 & 27 & 12064-2 & 4135 & 2014/05/17 & 2014/05/19 \\
 5 & 11416-2 & 1389 & 2012/02/11 & 2012/02/13
 & 28 & 12064-3 & 4135 & 2014/05/19 & 2014/05/21 \\
 6 & 11416-3 & 1389 & 2012/02/13 & 2012/02/15
 & 29 & 12240-1 & 4943 & 2014/12/15 & 2014/12/17 \\
 7 & 11673-1 & 2489 & 2013/02/17 & 2013/02/19
 & 30 & 12240-2 & 4943 & 2014/12/17 & 2014/12/19 \\
 8 & 11673-2 & 2489 & 2013/02/19 & 2013/02/21
 & 31 & 12240-3 & 4943 & 2014/12/19 & 2014/12/21 \\
 9 & 11673-3 & 2489 & 2013/02/21 & 2013/02/23
 & 32 & 12400-1 & 5865 & 2015/08/11 & 2015/08/13 \\
 10 & 11682-1 & 2501 & 2013/02/25 & 2013/02/27
 & 33 & 12400-2 & 5865 & 2015/08/13 & 2015/08/15 \\
 11 & 11682-2 & 2501 & 2013/02/27 & 2013/03/01
 & 34 & 12400-3 & 5865 & 2015/08/15 & 2015/08/17 \\
 12 & 11682-3 & 2501 & 2013/03/01 & 2013/03/03
 & 35 & 12528-1 & 6464 & 2016/04/04 & 2016/04/06 \\
 13 & 11710-1 & 2599 & 2013/03/30 & 2013/04/01
 & 36 & 12528-2 & 6464 & 2016/04/06 & 2016/04/08 \\
 14 & 11710-2 & 2599 & 2013/04/01 & 2013/04/03
 & 37 & 12663-1 & 7050 & 2017/06/16 & 2017/06/18 \\
 15 & 11710-3 & 2599 & 2013/04/03 & 2013/04/05
 & 38 & 12663-2 & 7050 & 2017/06/18 & 2017/06/20 \\
 16 & 11813-1 & 3049 & 2013/08/08 & 2013/08/10
 & 39 & 12664-1 & 7058 & 2017/06/22 & 2017/06/24 \\
 17 & 11813-2 & 3049 & 2013/08/10 & 2013/08/12
 & 40 & 12664-2 & 7058 & 2017/06/24 & 2017/06/26 \\
 18 & 11886-1 & 3326 & 2013/10/29 & 2013/10/31
 & 41 & 12664-3 & 7058 & 2017/06/26 & 2017/06/28 \\
 19 & 11886-2 & 3326 & 2013/10/31 & 2013/11/02
 & 42 & 12664-4 & 7058 & 2017/06/28 & 2017/06/30 \\
 20 & 11922-1 & 3490 & 2013/12/10 & 2013/12/12
 & 43 & 12719-1 & 7299 & 2018/08/20 & 2018/08/22 \\
 21 & 11922-2 & 3490 & 2013/12/12 & 2013/12/14
 & 44 & 12719-2 & 7299 & 2018/08/22 & 2018/08/24 \\
 22 & 11957-1 & 3648 & 2014/01/18 & 2014/01/20
 & 45 & 12719-3 & 7299 & 2018/08/24 & 2018/08/26 \\
 23 & 11957-2 & 3648 & 2014/01/20 & 2014/01/22
 & & & & & \\ [0.3em]
 \hline
\end{tabular}

\caption{Table listing the studied independently sampled CFARs and non-flaring ARs (NFARs) with their SHARP numbers and observation time windows; both $T_{start}$ and $T_{end}$ of the intervals correspond to 00:00 hr UTC}
\label{table:dataset2}
\end{centering}
\end{table}

\clearpage
\newpage

\section{Error Analysis} 
\label{app:Monte_Carlo}

\phantom{...}

Our approach of detecting PILs (see Section \ref{ssec:comp}) is sensitive to the 2D-configuration of the vertical magnetic field $B_z$, and thus, we assume that any uncertainty in the computed value of $L_{tot}$ is also a function of the overall configuration of the magnetic field. Since the noise level of the vertical photospheric images is $\sim$ 10 G \citep{Liu2012}, we performed a Monte Carlo simulation of the error in $L_{tot}$ (i.e. $\Delta L_{tot}$) by (i) adding a normally distributed error ($\mu$ = 0 G, $\sigma$ = 10 G) to the radial component of the magnetic field ($B_r$) for AR 11158 (15 February, 2011 at 00:00 UTC), (ii) subsequently extrapolating the data (PF) up to 3.24 Mm and then (iii) computing $L_{tot}$ at each height. This exercise was repeated for 100 iterations at each altitude and the results are tabulated below: \\

\begin{table}[ht]
\begin{center}
\hskip -3.0 cm
\begin{tabular}{| c | c | c | c | c | c | c | c |}
 \hline
 \multicolumn{8}{|c|}{} \\
 [-0.9em]
 \multicolumn{8}{|c|}{Monte Carlo estimates of $L_{tot}$ and $\Delta L_{tot}$ (100 iterations)} \\[0.4em]
 \hline
 \phantom{x} & \phantom{x} & \phantom{x} & \phantom{x} &
 \phantom{x} & \phantom{x} & \phantom{x} & \phantom{x} \\[-1.2em]
 No. & Height (Mm) & $L_{tot} \pm \Delta L_{tot}$ (Mm) & Uncertainty & No. & Height (Mm) & $L_{tot} \pm \Delta L_{tot}$ (Mm) & Uncertainty \\[0.2em]
 \hline
 1 & 0.00 & 226.4 $\pm$ 9.1 & 4.0\% & 6 & 1.80 & 74.04 $\pm$ 0.26 & 0.35\% \\
 \hline
 2 & 0.36 & 159.0 $\pm$ 4.6 & 2.9\% & 7 & 2.16 & 66.8 $\pm$ 3.2 & 4.8\% \\
 \hline
 3 & 0.72 & 131.45 $\pm$ 0.40 & 0.30\% & 8 & 2.52 & 62.1 $\pm$ 2.3 & 3.7\% \\
 \hline
 4 & 1.08 & 103.5 $\pm$ 2.3 & 2.2\% & 9 & 2.88 & 55.91 $\pm$ 0.006 & 0.011\% \\
 \hline
 5 & 1.44 & 79.27 $\pm$ 0.64 & 0.81\% & 10 & 3.24 & 55.41 $\pm$ 0.12 & 0.22\% \\
 \hline
\end{tabular}
\end{center}
\caption{Results of Monte Carlo simulation of $L_{tot}$ and $\Delta L_{tot}$ for AR 11158 on 15 February, 2011 at 00:00 UTC; $\Delta L_{tot}$ and the uncertainty are presented in figures with two significant digits}
\label{table:blank8}
\end{table}

The Monte Carlo analysis presented here (see Table \ref{table:blank8}) provides an estimate of the uncertainty in $L_{tot}$ i.e. $\Delta L_{tot}$, resulting from the observational uncertainty in the magnetic field. The propagation of errors through the extrapolation and application of the PIL detection algorithm appears to show no consistent pattern as a function of height. While the analysis is based on a single AR and therefore does not provide a statistical characterisation of the uncertainties across different ARs, the uncertainty in $B_r$ measurements suggests that the propagated uncertainties $\Delta L_{tot}$ are generally expected to remain well below 10\%. We therefore adopt 10\% as a conservative upper-bound estimate of the uncertainty associated with $L_{tot}$. Since $L_{tot}^*$ is defined as the maximum value of $L_{tot}$ over a 48-hr period, its uncertainty, $\Delta L_{tot}^*$ can also be assumed to be within 10\% of the $L_{tot}^*$ value. \\

\bibliography{sample631}{}

\bibliographystyle{aasjournal}

\end{document}